\documentclass[twocolumn,pra]{revtex4}
\usepackage{amsmath}
\usepackage{amssymb}
\usepackage{graphicx}

\begin{document}

\title{Towards quantum optics and entanglement\\
with electron spin ensembles in semiconductors}

\author{Caspar~H.~van~der~Wal}
\email[Corresponding author, e-mail: ]{c.h.van.der.wal@rug.nl}
\affiliation{Physics of Nanodevices Group, Zernike Institute for Advanced Materials,\\
University of Groningen, Nijenborgh 4, NL-9747AG  Groningen, The
Netherlands}
\author{Maksym~Sladkov}
\affiliation{Physics of Nanodevices Group, Zernike Institute for Advanced Materials,\\
University of Groningen, Nijenborgh 4, NL-9747AG  Groningen, The
Netherlands}

\date{\today}

\begin{abstract} 

We discuss a technique and a material system that enable the
controlled realization of quantum entanglement between spin-wave
modes of electron ensembles in two spatially separated pieces of
semiconductor material. The approach uses electron ensembles in GaAs
quantum wells that are located inside optical waveguides. Bringing
the electron ensembles in a quantum Hall state gives selection rules
for optical transitions across the gap that can selectively address
the two electron spin states. Long-lived superpositions of these
electron spin states can then be controlled with a pair of optical
fields that form a resonant Raman system. Entangled states of
spin-wave modes are prepared by applying quantum-optical measurement
techniques to optical signal pulses that result from Raman
transitions in the electron ensembles.







\end{abstract}

\maketitle


\section{\label{sec:introduction}Introduction}

Entanglement is the phenomenon that the quantum states of two (or
more) degrees of freedom are inseparable, and is arguably the most
distinct aspect of quantum theory \cite{nielsen2000book}. It results
in non-classical correlations between observable physical properties
of the two subsystems. For \textit{nonlocal entanglement} this
concerns two degrees of freedom that are spatially separated over a
large distance. The occurrence of such correlations has been
thoroughly tested in several experiments, and the results leave
little doubt that quantum theory provides the valid predictions.
Experimental realizations were, until now, carried out with pairs of
elementary particles or photons \cite{aspect1982prl,tittel1998prl},
or with spins in very simple quantum systems as for example trapped
ions \cite{rowe2001nature} or alkali atoms \cite{chou2005nature,
matsukevich2006prl}. It is nevertheless interesting to continue
research on the controlled realization of nonlocal entanglement with
other material systems, in particular with degrees of freedom in
solid state.

In part this interest is fundamental. Whether entangled states loose
their coherence in a different manner than superposition states of
individual degrees of freedom is still not fully understood
\cite{yu2004prl,dodd2004pra}. Recent developments here include an
all-optical experiment that showed that entanglement can be lost
much more rapidly than the loss of coherence in the two subsystems
\cite{almeida2007science}. Another interesting result from work with
entangled photon pairs showed that the relation between the amount
of entanglement and the degree of mixedness of a two-particle state
can only be represented by a plane of possibilities. Specific points
in this plane depend on the nature of the environment that is
decohering the initial maximally-entangled pure state
\cite{puentes2007,aiello2007pra}. Furthermore, it is still not
firmly established that quantum theory does not break down when
applied to collective or macroscopic degrees of freedom
\cite{leggett1985prl,leggett1998book}. This justifies a study of how
entangled states can be realized in solid state, and how these
states loose their coherence: solid state can provide model systems
with complex (collective) degrees of freedom, or systems with
elementary degrees of freedom in a complex environment.

Research on the controlled realization of entanglement in solid
state systems is also driven by the prospect that it may provide
tools for quantum information technologies. Relevant to the
discussion here is a proposal for long-distance quantum
communication \cite{duan2001nature}, that was until now mainly
explored with ensembles of alkali atoms
\cite{chou2005nature,matsukevich2006prl}, or alkali-atom-like
impurities in solids (Refs.~\cite{longdell2005prl,rippe2005pra,
guillot2007jlum,louchet2007jlum,dutt2007science} and related
articles in this issue). However, widespread implementation favors a
technique that can be implemented in micron-scale devices that fit
inside optical fibers, which are compatible with high-speed
opto-electronic operation \cite{lukin2001nature}. Here, the
electronic and optical properties of III-V semiconductors outperform
the atomic or impurity-based systems. The coherence times of degrees
of freedom in these materials, however, tend to be too short for any
realization of quantum information technology in the near future,
but are long enough for initial experimental studies on entangled
states.

We discuss here a technique that enables the controlled realization
of nonlocal entanglement between spin-wave modes in ensembles of
conduction-band electrons, which are located in two spatially
separated pieces of GaAs semiconductor material. We also outline the
material properties of a GaAs quantum well system where this
technique can be implemented. In Section~\ref{sec:entangle} we
discuss an approach where quantum-optical measurement techniques are
used for preparing entangled states of spin degrees of freedom in
ensembles of three-level quantum systems. Subsequently, in
Section~\ref{sec:gaasmatter}, we present a GaAs heterostructure
material that is suited for realizing such an ensemble of
three-level quantum systems.


\section{\label{sec:entangle}Preparing and detecting entangled states via quantum optical measurement}

We propose here to use the so-called DLCZ scheme
\cite{duan2001nature} for preparing nonlocal entanglement with solid
state devices. The main idea behind this approach is that
spontaneous emission of a quantum optical pulse results in quantum
correlations (entanglement) between the state of the optical pulse
and the state of the system that emits. To illustrate this, consider
a two-level system that is initially in its excited state $\mid
\uparrow \rangle$. It is emitting a single photon while relaxing to
its ground state $\mid \downarrow \rangle$. If we would be able to
have control over this process such that it relaxes to a
superposition of the states $\mid \uparrow \rangle$ and $\mid
\downarrow \rangle$, the system would emit an optical pulse that is
a superposition of the states with 0 and 1 photon, $|0_{puls}
\rangle$ and $|1_{puls} \rangle$. The quantum state of the system
and the optical pulse are then in fact entangled, and the only pure
states that can describe the state of the combined system are of the
form $| \Psi_{com} \rangle = c_{\uparrow}\mid \uparrow \rangle
|0_{puls} \rangle + c_{\downarrow}\mid \downarrow \rangle |1_{puls}
\rangle$.

\begin{figure}[t!]
  \includegraphics[width=85mm]{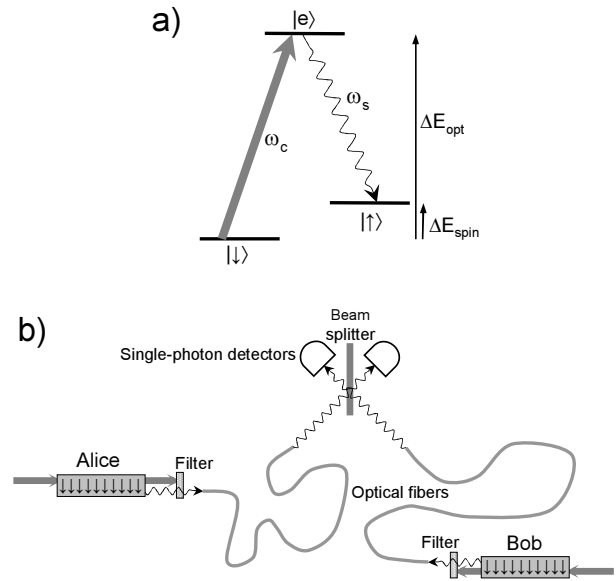}
  \caption{
  \textbf{a)} A three-level system with $\Delta E_{spin}<<\Delta E_{opt}$.
  The transition between two low-energy spin states
  $\mid  \downarrow \rangle$ and $\mid  \uparrow \rangle$ under spontaneous emission
  of a signal photon from the transition $|e \rangle - \mid \uparrow \rangle$
  (with energy $\hbar \omega_{s}$) can be controlled with
  an optical field (tuned to photon energy $\hbar \omega_{c}$)
  driving the transition $\mid \downarrow \rangle - |e \rangle$.
  The two legs can be selectively addressed using the optical frequency
  difference or their dependence on the polarization of the fields.
  \textbf{b)} Scheme for entangling the states of spin-wave modes
  in two different spin ensembles, see the text for details.}
  \label{fig:entangle}
\end{figure}

Such control over spontaneous emission can be realized with a
three-level Raman system (Fig.~\ref{fig:entangle}a). When this
system is initially in the state $\mid \downarrow \rangle$, there
will be only spontaneous emission of a Raman photon from the
transition $|e \rangle - \mid \uparrow \rangle $ while a control
field is driving the $\mid \downarrow \rangle - |e \rangle$
transition. Figure~\ref{fig:entangle}b illustrates how an extension
of this scheme can be used to entangle the states of two three-level
systems that are at different locations. Say Alice and Bob both have
an identical version of such a three-level system prepared in the
state $\mid \downarrow \rangle$. They both use a classical field to
drive the $\mid \downarrow \rangle - |e \rangle$ transition, in
order to get very weak spontaneous emission from the $|e \rangle -
\mid \uparrow \rangle $ transition, such that each system emits an
optical pulse that is a superposition of the photon-number states
$|0_{puls} \rangle $ and $|1_{puls} \rangle $ (note that each of
these pulses is then entangled with the system that emitted it). The
timing and propagation of these two pulses should be controlled such
that they arrive at the same time at a measurement station, that
consists of a 50/50 beam splitter with a photon counter at each of
its two output channels. If the number of photons in the pulses are
now measured after combining the two pulses on the beam splitter,
there is some probability that one of the two detectors counts 1
photon and the other 0 photons. In that case, the total number of
spin flips in the two three-level systems is 1, but it is impossible
to tell which of the two emitted the photon. As a result, the
systems of Alice and Bob have been projected onto an entangled state
of the form $| \Psi_{AB} \rangle = \frac{1}{\sqrt{2}} \left( \mid
\uparrow_{A} \rangle \mid \downarrow_{B} \rangle + {\rm e}^{i
\varphi} \mid \downarrow_{A} \rangle \mid \uparrow_{B} \rangle
\right)$ (where the phase $\varphi$ can be derived from experimental
conditions \cite{cabrillo1999pra,duan2001nature}).

Figure~\ref{fig:entangle}b depicts in fact emission from ensembles
of three level systems. For weak (slightly detuned) driving of the
$\mid \downarrow \rangle - |e \rangle$ transition, the expectation
value for the total number of $|e \rangle - \mid \uparrow \rangle $
photons emitted by an ensemble of identical three-level systems can
still be less than 1 photon. Notably, the spin excitation is then
not stored on an individual three-level system. Instead, it is
stored as a spin-wave mode (collective spin excitation) in this
medium, with each three-level system having its spin flipped only by
a very small amount. Thus, one can also use this approach for
preparing entanglement between spin-wave modes in two different
ensembles.

These ensembles should have a long elongated shape that is co-linear
with the driving field. An important advantage of using such
ensembles is that spontaneous emission becomes highly directional
\cite{duan2001nature}, with emission predominantly co-propagating
with the driving field. In principle the system will emit very weak
in all directions, but an initial spontaneous emission event
(extremely weak, far below the single-photon level) is strongly
amplified (gain) when it co-propagates with the driving field
\cite{vanderwal2003science}. For very weak driving, the total energy
in all of the spontaneous emission can still be at the single-photon
level, and the gain then ensures that emission into the desired
direction is exponentially stronger than into other directions.
Thus, the collection efficiency for the total number of emitted
photons by such ensembles can be near unity. This removes the need
for using high-finesse optical cavities as in cavity-QED
experiments, which is technically very demanding
\cite{mckeever2004science,hennessy2007nature}.


One should also be able to confirm that entangled states have been
prepared by reading out the states of each ensemble of a pair that
has been entangled. Correlations between the spin excitations in the
two ensembles (Fig.~\ref{fig:entangle}b) can be studied with an
optical readout scheme that uses the inverse of the initial Raman
transition. For each system separately, the number of flipped spins
in its ensemble can be measured using a control field that is now
driving the $ \mid \uparrow \rangle - |e \rangle$ transition. This
converts the spin state that is stored in an ensemble into the state
of an highly-directional optical pulse (again a superposition of
photon-number states $|0_{puls} \rangle $ and $|1_{puls} \rangle $)
that results from a subsequent $|e \rangle - \mid \downarrow \rangle
$ transition. This process fully returns the spin excitation into
the $\mid \downarrow \rangle $ state. The detection should now
directly count the number of photons in the emission from the
ensemble that is measured (not using a configuration with a beam
splitter). Each of the two ensembles should be measured separately
in this manner. If the two ensembles were prepared in a state of the
form $| \Psi_{AB} \rangle = \frac{1}{\sqrt{2}} \left( \mid
\uparrow_{A} \rangle \mid \downarrow_{B} \rangle + {\rm e}^{i
\varphi} \mid \downarrow_{A} \rangle \mid \uparrow_{B} \rangle
\right)$, the number of detected photons from the ensemble of Alice
can be 0 or 1, each with probability $\frac{1}{2}$. However, for
either measurement outcome, subsequent measurement of the number of
photons emitted by Bob's ensemble must yield that it is perfectly
anti-correlated with the result of Alice.

Such measurements can already provide evidence for the quantum
nature of these correlations (in particular, the variance of these
photon-count correlations should show strong sub-Poissonian
statistics \cite{vanderwal2003science}). However, it does not yet
allow for a formal test of Bell inequalities (testing for
non-classical correlations), since this requires the ability to
rotate the basis in which the state of each of the two-level systems
is measured (with respect to the basis defined by $|0_{puls} \rangle
$ and $|1_{puls} \rangle $). This cannot be performed directly with
a readout technique based on photon-number measurements. To overcome
this, the observation of entanglement between two ensembles of
alkali atoms \cite{chou2005nature} used an approach where a local
phase shift was applied to one of the two systems, either to the
optical signal pulse from readout \cite{chou2005nature} or to the
stored spin excitation. However, the readout then requires once more
to combine the signal pulses from readout of the two ensembles on a
beam splitter, and to study the interference fringe that results
from the local phase shift. A scheme that only relies on local
readout of each ensemble can be realized when the states of both
Alice and Bob are not stored in a single ensemble but in a pair of
ensembles \cite{duan2001nature}. The photon-number readout can then
be implemented with a certain setting for a phase difference between
the states of these two ensembles. However, both of these approaches
require that the path length between the ensembles and detector
stations are stabilized with interferometric precision. An
alternative more robust approach could be realized with alkali atom
ensembles \cite{matsukevich2006prl} and used the fact that in these
systems the states $ \mid \downarrow \rangle$  and
$ \mid \uparrow \rangle$
consist of multiple (degenerate) Zeeman sublevels. How a
spin excitation is distributed over these Zeeman sublevels is then
mapped onto two orthogonal polarizations of a signal field, and
polarization selective readout then enables to rotate the basis in
which signal fields are measured. Other solutions that are
technically even less demanding are currently investigated
\cite{lan2007prl,zhao2007prl,jiang2007pra,chen2007prl}.

Applying this quantum-optical measurement scheme for preparing
entangled states in spatially separated electronic devices is an
interesting alternative to related research that uses electronic
control and measurement techniques. Activities here use for example
electron spins in quantum Hall states \cite{beenakker2003prl} or
quantum dots \cite{petta2005science}, or superconducting qubits
\cite{plantenberg2007nature}. A first advantage of this quantum
optical approach is that it naturally allows for having the two
devices separated by a large distance, whereas for electronic
control coherent interactions are typically limited to short
distances. More importantly, it allows one to use photon-number
detection. This is a unique quantum measurement tool in the sense
that projective measurement can be used for preparing states with
very high fidelity. Tools for electronic readout have typically much
higher noise levels, which results in a much weaker correlation
between a measurement outcome and the state of the quantum system
immediately after measurement.


\section{\label{sec:gaasmatter}$\textrm{GaAs}$ quantum wells as a medium for quantum optics}

We now discuss how such an ensemble of three-level systems can be
implemented in a GaAs quantum well system. The techniques presented
in the previous section have been mainly developed and explored with
ensembles of alkali atoms
\cite{matsukevich2006prl,chou2005nature,lukin2003rmp}, and
developing a realization in solid state can thus be viewed as an
attempt to mimic an ensemble of alkali atoms. The key properties of
the three-level system as in Fig.~\ref{fig:entangle}a are then that
the splitting $\Delta E_{spin}$ is homogeneous for an ensemble, and
that superpositions of $\mid \downarrow \rangle$ and $\mid \uparrow
\rangle$ have a long coherence time $T_{\downarrow \uparrow}$.
Further, these two states $\mid \downarrow \rangle$ and $\mid
\uparrow \rangle$ must both have a strong optical transition to a
common excited state $|e \rangle$ that can be addressed selectively.
For each leg the spontaneous emission life time must be much shorter
than $T_{\downarrow \uparrow}$, a requirement that overlaps with
conditions for electromagnetically induced transparency
\cite{fleischhauer2005rmp}. Transient signatures of
electromagnetically induced transparency were already observed in
related work using excitons in undoped GaAs quantum wells
(Ref.~\cite{palinginis2005apl} and references therein).

\begin{figure}[t!]
  \includegraphics[width=85mm]{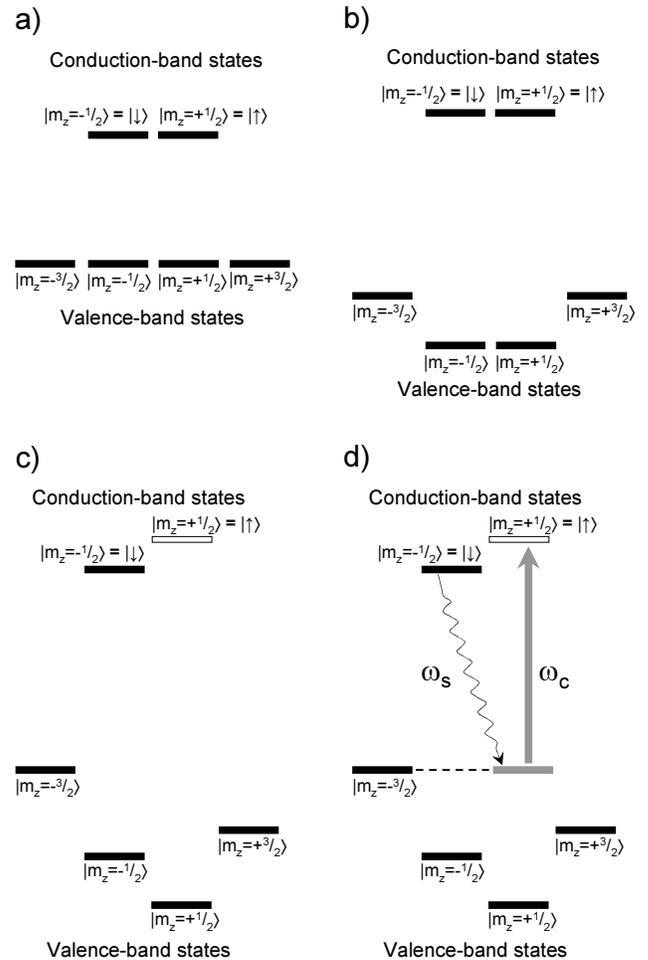}
  \caption{
  \textbf{a)} The lowest conduction-band states and highest valence-band states
  (for electron wavevector $k=0$) for bulk $n$-GaAs.
  This system has good selection rules for optical transitions
  between energy states with well-defined
  angular momentum in $z$ direction (quantum number $m_z$).
  \textbf{b)} As panel a), but now for a $n$-doped GaAs quantum well
  system (in the $x$-$y$ plane). Quantum well confinement lifts the
  degeneracy between the heavy-hole ($m_z=\pm \frac{3}{2}$) and
  light-hole ($m_z=\pm \frac{1}{2}$) states
  (levels for in-plane electron wavevector $k=0$).
  \textbf{c)} As panel b), but now for this system in a
  magnetic field in the $-z$-direction,
  bringing the system in the $\nu = 1$ quantum Hall state.
  Now, the levels indicate the spin-polarized
  Landau levels, and are all filled (thick solid line), except for the
  conduction-band $|m_z=+ \frac{1}{2} \rangle$ state (open line) which is fully empty.
  The energy levels represent here electronic cyclotron states (dispersion free).
  \textbf{d)} As panel c), but now displaying the effect of hole mixing
  (here only sketched for the highest hole level in the valance band,
  which has both $|m_z=- \frac{3}{2} \rangle$  and $|m_z=+ \frac{1}{2} \rangle$ character).
  Due to this hole mixing, the transitions labeled with
  $\omega_c$ and  $\omega_s$ can be selectively
  addressed with optical fields that propagate in plane and
  have orthogonal linear polarizations.}
  \label{fig:selrules}
\end{figure}

We propose here to realize an ensemble of three-level systems with
an $n$-doped GaAs quantum well system, building on seminal work by
Imamoglu \cite{imamoglu2000optcom}. These $n$-doped GaAs materials
combine relatively long coherence times for electron spin
superposition states ($\gtrsim 10\;{\rm ns}$
\cite{kikkawa1998prl,teraoka2004physE,gerlovin2007prb}), with strong
optical transitions across the gap that obey good selection rules
\cite{weisbuch1991book}. Transitions between the highest valence
band states and lowest conduction band states follow the selection
rules for transitions between a $p_{3/2}$ and $s_{1/2}$ manifold
\cite{cohent1977book}, as represented for bulk GaAs in
Fig.~\ref{fig:selrules}a. The continuous density of states for bulk
GaAs develops into an atom-like discrete set of levels when a
two-dimensional electron system (Fig.~\ref{fig:selrules}b) is
brought into the quantum Hall regime by applying a strong magnetic
field (Fig.~\ref{fig:selrules}c), because the electrons then
condense into cyclotron orbits. The number of these so-called Landau
levels that are filled at a certain magnetic field defines the
filling factor $\nu$, where each spin-resolved Landau level is
counted individually (it can be shown that $\nu$ equals the ratio of
areal electron density and areal density of flux quantums
\cite{weisbuch1991book}). We consider the quantum Hall state at
filling factor $\nu = 1$, where the conduction-band Landau levels
are fully spin polarized. At $\nu = 1$, the lowest Landau level
(further denoted as $\mid \downarrow \rangle$) is fully occupied,
while all higher Landau levels are fully unoccupied. The density of
states for the valance band is now also a discrete set of Landau
levels that are all fully filled. Note that we assume that the
magnetic field is in the $-z$-direction, in order to have
Fig.~\ref{fig:selrules} compatible with Fig.~\ref{fig:entangle}a (in
GaAs the electron g-factor is negative).

The system as sketched in Fig.~\ref{fig:selrules}c allows for
implementing a three-level system, with the lowest spin-up and
spin-down Landau levels in the conduction band serving as the two
low-energy spin states $\mid \downarrow \rangle$ and $\mid \uparrow
\rangle$. These both have, for example, an optical transition to the
valance-band $|m_z=-\frac{1}{2} \rangle$ state, as the selection
rules only allow for transitions with $\Delta m_z \in \{ -1, 0,
+1\}$. For GaAs quantum well systems, selective addressing of these
two optical transitions must rely on polarization selection rules,
since the Zeeman splitting for the conduction-band electrons is not
in excess of the narrowest line width that can be obtained for
optical transitions. In practice, the narrowest lines ($\approx 0.2
\; {\rm meV}$) can be obtained for a quantum well width of $\approx
20 \; \rm{nm}$ \cite{langbeinPRB2000,eshlagi2000thesis,schwedt2003physstatsol},
while the electron g-factor $g_e \approx -0.4$
\cite{hannak1995solstatcom}. This gives an electron Zeeman splitting
of $\approx 0.25 \; {\rm meV}$ in a field of 10 Tesla (the electron
g-factor is not enhanced for spin-wave modes with a wavelength
longer than the magnetic length \cite{pinczuk1992prl}, as observed
in electron spin resonance studies on such ensembles
\cite{stein1983prl,teraoka2004physE}). The other energy splittings
(partly shown in Fig.~\ref{fig:selrules}c) are all larger and allow
for using spectral selectivity. These splittings are further
discussed below.


For the quantum well width that we choose to consider here ($\approx
20 \; \rm{nm}$) there are hole-mixing effects that cannot be
neglected: the hole energy levels are then a superposition of two or
more different angular momentum states (each characterized by a
quantum number $m_z$). These effects, however, can be used for
implementing a more convenient control scheme, where the three-level
system is formed by $\mid \downarrow \rangle$, $\mid \uparrow
\rangle$ and the highest Landau level in the valence band. This
means that one can work with the first optical transition that becomes
available when increasing the photon energy from within the gap, and
this transition also has the narrowest line. In practice this an
important advantage. Using this transition also has the advantage
that it is the most isolated level: lower Landau levels in the
valance band for holes with (predominantly) $m_z = \pm \frac{1}{2}$
character may be very close to levels for holes with (predominantly)
$m_z = \pm \frac{3}{2}$ character for which the quantum number for
confinement in the well or Landau level orbital is increased by one
\cite{ancilotto1988prb} (levels not shown in
Fig.~\ref{fig:selrules}c). How such a three-level system can be
implemented in a model system where hole-mixing is included is
depicted in Fig.~\ref{fig:selrules}d. For the highest valence-band
Landau level, the heavy-hole $|m_z=-\frac{3}{2} \rangle $ state
mixes predominantly with the light-hole $|m_z=+\frac{1}{2} \rangle $
state \cite{ancilotto1988prb}. In principle, the $|m_z=-\frac{3}{2}
\rangle $ state also mixes with the $|m_z=-\frac{1}{2} \rangle $
state, but this mixing is much weaker. Moreover, its contribution to
the optical transition is negligible (parity forbidden) if the
quantum well is symmetric, since the mixing is in fact with a
$|m_z=-\frac{1}{2} \rangle $ state of a different orbital (it has
the quantum number for quantum-well confinement increased by one).

\begin{figure}[t!]
  \includegraphics[width=85mm]{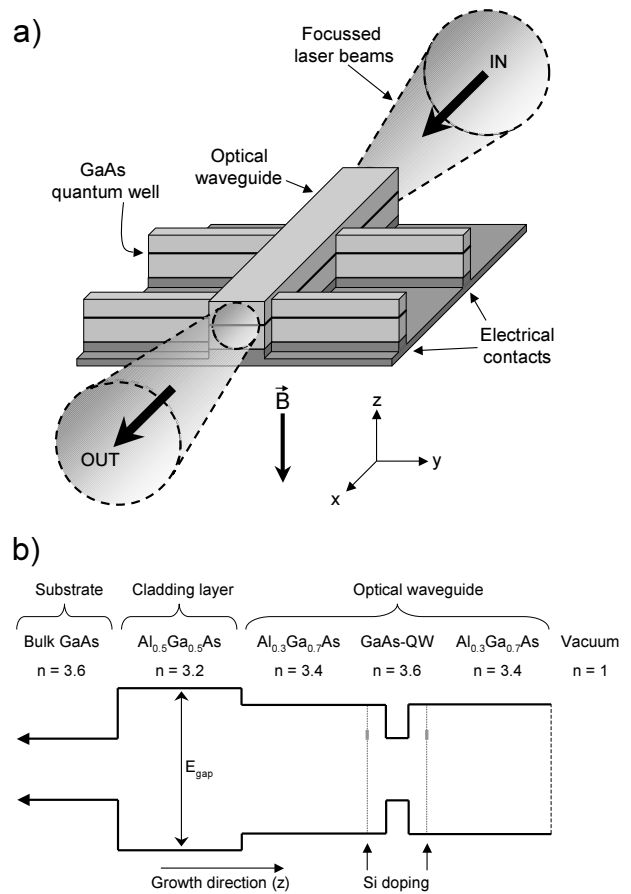}
  \caption{
  \textbf{a)} Optical waveguide with an electron-spin ensemble in
  a GaAs quantum-well (not to scale), etched out of a
  GaAs/${\rm Al}_{x}{\rm Ga}_{1-x}{\rm As}$ heterostructure.
  The thick solid black line represents the quantum well layer and
  forms a central layer all through the waveguide.
  In an external magnetic field the optical
  excitation spectrum for electrons is as in Fig.~\ref{fig:selrules}d
  (for being compatible with Fig.~\ref{fig:selrules} the field is in the
  $-z$-direction).
  Optical control and signal fields co-propagate through the waveguide.
  Electrical contacts are used for in-situ monitoring of the quantum Hall
  effect.
  \textbf{b)} Design of the wafer material. It can be grown with conventional techniques
  for epitaxial growth of GaAs/${\rm Al}_{x}{\rm Ga}_{1-x}{\rm As}$ heterostructures. The figure shows
  the energy gap profile (not to scale, and neglecting band-bending effects near
  hetero-interfaces) and lists the index of refraction $n$ along the
  growth direction ($z$).
  We use that for GaAs/${\rm Al}_{x}{\rm Ga}_{1-x}{\rm As}$ material the energy gap increases
  with increasing Al content, while the index of
  refraction decreases with increasing Al content.}
  \label{fig:waveguide}
\end{figure}

Figure~\ref{fig:selrules}d depicts the optical transitions that can
be used when this highest hole level is used for operating a
three-level system. Note that we consider here co-propagating
control field $\omega_c$ and signal field $\omega_s$ in the plane of
the quantum well (in $x$-direction, see also
Fig.~\ref{fig:waveguide}a), since this allows us to work with long
elongated spin ensembles. This also gives convenient polarization
selection rules. The control field $\omega_c$ should address a
$\Delta m_z=0$ transition, which couple to fields that are linearly
polarized along the $z$-direction. The signal field $\omega_{s}$
concerns $\Delta m_z= \pm 1$ transitions, which couples for the case
of \textit{in-plane} emission to fields with an orthogonal linear
polarization (along the $y$-direction, it would be circular
polarized for propagation orthogonal to plane
\cite{weisbuch1991book,cohent1977book}). This holds both for the
path $\mid \downarrow \rangle$-$|m_z=-\frac{3}{2} \rangle $ and (as
sketched) the path $\mid \downarrow \rangle$-$|m_z=+\frac{1}{2}
\rangle $. The analogy with Fig.~\ref{fig:entangle}a is more evident
when describing the process in terms of holes
\cite{imamoglu2000optcom}. The control field $\omega_c$ is then
driving a hole from the fully filled conduction band level $\mid
\uparrow \rangle$ to the highest valence-band level. This hole can
then relax to the $\mid \downarrow \rangle$ level by emission into
the $\omega_s$ field (for this picture the arrows in
Fig.~\ref{fig:selrules}d should point in the opposite direction).
Note, however, that the relevant coherence time for the two
low-energy states $\mid \uparrow \rangle$ and $\mid \downarrow
\rangle$ is nevertheless the spin coherence time for electrons in
the conduction band.

Figure~\ref{fig:waveguide}a presents a device structure that can
realize this model system. The GaAs quantum well is embedded in an
${\rm Al}_{x}{\rm Ga}_{1-x}{\rm As}$ single-mode optical waveguide.
Narrow (as compared to the optical wavelength) electrical contacts
on the side of this waveguide serve for in-situ monitoring of the
quantum Hall state of the electron ensemble inside the waveguide.
The quantum well inside the piece of waveguide then contains a
single electron ensemble, such that the device structure of
Fig.~\ref{fig:waveguide}a represents one of the two spin ensembles
in the scheme of Fig.~\ref{fig:entangle}b. This device structure
naturally implements the situation that the control and signal
fields co-propagate and have perfect overlap with a spin ensemble
that has a long elongated shape. Furthermore, the engineering of the
quantum well and the waveguide with cladding layer naturally fit
together, using the fact that for ${\rm Al}_{x}{\rm Ga}_{1-x}{\rm
As}$ material the energy gap increases with increasing Al content,
while the index of refraction decreases with increasing Al content
\cite{saleh1991book}. Figure~\ref{fig:waveguide}b presents how this
can be realized. The GaAs quantum well is located in the middle of
an ${\rm Al}_{0.3}{\rm Ga}_{0.7}{\rm As}$ layer that serves as the
core of the waveguide. This waveguide core is transparent for the
optical fields that are used since the gap of ${\rm Al}_{0.3}{\rm
Ga}_{0.7}{\rm As}$ is larger than the gap of GaAs. The ${\rm
Al}_{0.3}{\rm Ga}_{0.7}{\rm As}$ layer acts as a waveguide core
since it is embedded between layers with a lower index of refraction
(either formed by ${\rm Al}_{0.5}{\rm Ga}_{0.5}{\rm As}$ or vacuum).

We end this section with further quantifying the material
parameters. We already argued that a symmetric GaAs quantum well
system with a width of 20 nm is the optimal choice. To bring it in
the quantum Hall state $\nu=1$ with a magnetic field of about 10~T,
the quantum well should contain a high-mobility electron gas with a
density of $n_s \approx 2.4 \cdot 10^{15} \; {\rm m}^{-2}$. In a
field of 10~T, the Landau levels are as depicted in
Fig.~\ref{fig:selrules}d, and we will discuss here the other energy
splittings of this system. The literature is not very conclusive
about the effective g-factor for the light- and heavy-hole levels.
This results from the fact that these depend on the quantum well
width, Al content of the ${\rm Al}_{x}{\rm Ga}_{1-x}{\rm As}$
barriers, strength and direction of the magnetic field, and hole
mixing effects. Nevertheless, most results indicate that the Zeeman
splittings for holes at 10~T in a 20 nm system are substantially
larger than $\approx 0.2 \; {\rm meV}$
\cite{ancilotto1988prb,traynor1997prb}. Typical values are close to
the energy spacings to the next hole Landau levels (with identical
value for $m_z$, but with the Landau-orbital quantum number one
higher). These splittings are under these conditions all $\gtrsim 3
\; {\rm meV}$, while the splittings between conduction-band Landau
levels are much larger thanks to the low effective mass of electrons
\cite{ancilotto1988prb}. Finally, the energy spacings between the
subbands due to confinement are for this system $\approx 25 \; {\rm
meV}$ for electrons, $\approx 5 \; {\rm meV}$ for heavy holes and
$\approx 20 \; {\rm meV}$ for light holes
\cite{weisbuch1991book,eshlagi2000thesis}. This also sets the scale
for the splitting between the highest heavy- and light-hole levels
(Fig.~\ref{fig:selrules}b), which is about $\approx 5 \; {\rm meV}$
\cite{weisbuch1991book}.


\section{\label{sec:conclusions}Conclusions}

The reasonably long coherence times for electron spin ensembles in
$n$-doped GaAs materials allows for studies of how such ensembles
can act as a medium for quantum optics. We showed that this idea is
feasible, and that this allows for preparing entanglement between
states of spin wave modes in two different ensembles. For initial
studies, an $n$-doped GaAs quantum well system in the quantum Hall
$\nu=1$ state provides the most promising model system. The electron
ensembles are addressed by placing the quantum wells inside optical
waveguides, with in-plane propagation of optical control and signal
fields. Realizing such systems is compatible with standard epitaxial
growth techniques for GaAs/${\rm Al}_{x}{\rm Ga}_{1-x}{\rm As}$
heterostructures. We analyzed that an optimal system is formed by a
symmetric GaAs quantum well of about 20 nm width. In this system one
can address electron spin degrees of freedom inside ensembles of
three-level quantum systems with optical transitions across the gap.
The most suitable three level system uses transitions between the
conduction band spin states and the highest Landau level of the
valence band. Selective control over these two transitions is
possible with polarization selection rules and using hole-mixing
effects that naturally occur in this system.

Progress towards the realization of entanglement with such a system
first requires spectroscopy with fields that propagate in plane to
confirm the optical selection rules (in particular with respect to
the hole mixing). Also Pauli blocking when driving a completely
filled Landau level needs to be demonstrated. A crucial next step is
then to demonstrate electromagnetically induced transparency (EIT)
\cite{fleischhauer2005rmp}, as this provides evidence that a medium
is suited for the quantum optical techniques that we discussed here.
If these steps are successful, this clean material system is a very
promising candidate for studies of entanglement with ensembles of
conduction band electrons in solid state. In particular, the
observed long spin coherence times for electron spin ensembles imply
that the Zeeman splittings are very homogeneous in these ensembles.
This allows to generate Raman scattered fields from two different
ensembles that are centered at identical optical frequencies, while
their spectral width is tuned by the EIT bandwidth \cite{fleischhauer2005rmp}.
Consequently, the two signal pulses then have very good spectral overlap, and preparing
entanglement by interfering these two pulses on a beam splitter
should indeed be possible.

\begin{acknowledgments}
We thank X.~Liu, D.~Reuter, D.~Gerace, H.~E.~Tureci and A.~Imamoglu
for help and stimulating discussions. This research is supported by
the Netherlands Organization for Scientific Research (NWO).
\end{acknowledgments}


\end{document}